\documentclass[11pt]{article}
\usepackage[T2A]{fontenc}
\usepackage[cp866]{inputenc}
\usepackage[english]{babel}
\usepackage[dvips]{graphicx }
\usepackage{amsmath,euscript,amssymb}
\newcommand{\be}{\begin{equation}}
\newcommand{\ee}{\end{equation}}
\newcommand{\bea}{\begin{eqnarray}}
\newcommand{\eea}{\end{eqnarray}}
\def\vh{\varphi}

\title{Creation of the Universe}

\author{ Victor N. Pervushin,\\
Bogoliubov Laboratory of Theoretical Physics,\\
Joint Institute for Nuclear Research, 141980 Dubna, Russia}

\begin{document}

\maketitle

\begin{abstract}

 Possibilities for solution of the problem of creation of the Universe
  from a physical vacuum
 in the framework the General Relativity
  and  modern quantum field theory are discussed
   in the context of the official doctrine accepted  in  Trinity
   College at the Newton time.

\end{abstract}

\vspace{2cm}

\hfil {\tiny A Christians in  Science Conference
 {\small ``Design and Purpose in the Universe''}

\hfil Saturday 18th March 2006, St. Nicholas' Church, Market
Place, Durham, DH1 1JP} { "www.cis.org.uk"}

\vspace{1cm}

\tableofcontents

\newpage

\section{Physical ``Laws'' and  ``Data'' }

 The founder of the first physical theory Isaac Newton
 gave to it some pattern of the official doctrine accepted in
 his Trinity College.
 Beginning with  Newton's mechanics all physical theories  remember
 to a some extent the Christian theory of
  a human soul being in two {\it kingdoms}\,:~ a {\it kingdom} of
{\it laws} and a {\it kingdom} of {\it wills}.

 In any physical theory, a {\it kingdom of laws} is a set of equations of
 motion obtained by varying an {\it action}.
 A {\it kingdom of wills} is associated with a set of initial data
 which
 Laplace  still required for unambiguous solutions of the Newton
 equations in order ``to explain the World without God''.

 The initial data of an object  are measured
 by a set of physical instruments
 and are given by a {\it will} of
 experimenters.
  {\it Laws} of nature do not depend on the initial
 data but  results of application of these {\it laws} can depend.
 The Newton equation of a train does not depend on the initial data, i.e. on
the {\it will} of a passenger of this train who chose  his initial
position
 and the speed of the train, but the final
result of the solution of this equation  is a consequence of both
the {\it will} of the passenger and  the {\it law} of nature.

 In contrast to mathematician Laplace, physicist Newton understood
  that in any explanation
 of the Design and Purpose in the Universe  physicists
 should take into account both the  {\it kingdoms} of {laws}
 and  {``data''} given by God in a comoving reference
 frame   where  the Universe was created.

    This comoving frame is identified by modern observers
   with the
 Cosmic Microwave Background (CMB) radiation
 distinguished by its dipole component measured \cite{WMAP} in the frame
 of an Earth observer. %Who chose this frame at the time
% of formation of CMB when an Earth observer was only
% in the Project?

 Formulations of Eistein's General
 Relativity  and quantum field theory
  in a concrete frame were
 fulfilled  by the founders of modern physics
 (for details
 see \cite{dir} -- \cite{pol})\footnote{
  I was an
eye-witnesser of
  that at the beginning of the 1970s theoreticians
 restricted
 quantum field theory
  by  describing mainly scattering
 processes that do not depend on the data together with
 their reference frames.
 This restriction allowed to formulate
 the frame free Faddeev -- Popov method \cite{fp1} that gave an essential
 simplification
 of the proof of renormalizabillity (finiteness) of the Standard Model
 marked by the Nobel Prize to t'Hooft and Veltman \cite{thw}.
 But this restriction   lost
 the historical pathway of
  reference frames  in physical theories beginning with
 Copernicus until Dirac \cite{Dir} and Schwinger \cite{sch2}.
 }. Now I present here the results of our attempts
 with my collaborators \cite{pvn7} --
 \cite{pvng8b}
 to use these formulations, in order
  to describe  the creation of the Universe
 in the framework of modern physical theory in agreement with
 observational data. %, by analogy with the
% description of a moving particle in the Newton mechanics?

\section{Action and Interval as ``Foundations of Physics''}

  The official doctrine accepted in
  Trinity College
  {\it ``In the beginning was the Word ..., and the
 Word was  God''} (John 1:1) can be  consistent with the modern physics,
 if  God can be visible due to
 His action.
 The  ``action'' of the unified theory was restored
  by  both experimentalists  and
 theoreticians  in the 19th and 20th centuries in the context
 of the field nature of matter and space-time in the form of  a sum of
  Hilbert's  action of the Einstein General Relativity \cite{H}
  and the action of
 the Weinberg - Salam - Glashow Standard Model
  of elementary particles \cite{db}
   settled in the Riemannian space-time defined by a geometric interval.
   Scientists believe in that
   equations of motion
 obtained by varying this action and supplemented by the initial data
  in the CMB  frame can  explain the
 origin of the Universe.
 %What does a ``concrete frame'' mean in the modern theories?

%``Make His paths straight.'' (Mark 1:3)

\section{A  Frame for Constrained {``Data''}}
   In General Relativity (GR)
   and quantum field theories, a comoving frame was introduced  by
  Dirac, Heisenberg and Pauli, Fermi, Schwinger, and
 other physicists until the 1960s
 on the level of the action (see \cite{dir} -- \cite{pol})
  because
 it was the most  straight path to determine the
 spectrum of the elementary and collective excitations and
 to separate equations of motion from a set of
 {\it constraints} between the initial data. %\footnote{
 %In the next years, when
% the frame dependence of the physical results were considered
% as nonessential,
% frames were established on the level of equations of motion.
 %Really, the frame-equation path does not differ
% from the ``straight'' frame-action one, if we restrict tasks of
% theoretical physics by  description of scattering processes
% of elementary particles in the framework of the standard perturbation
% theory \cite{fp1}.
%I am convinced that the ``straight'' frame-action  path leads to
%results
% that essentially differ from  the frame-equation path ones, if we describe
%  collective phenomena of the type of
% bound states \cite{pvn4,pvn7},
% the string spectrum, and
%  cosmic evolution in GR \cite{pvng7a}.}.
 These {\it constraints} of the data in all modern relativistic theories are principally
 new elements in comparison with the Newton mechanics.

 A contemporary colleague of Newton in Trinity College
  could  find that the constrained relativistic systems
 led to the official Trinity doctrine
 rejected by Isaac
 Newton\footnote{Introducing the standards
 of consistent
mathematical proof into physics in terms of clear absolute notions
 Newton was looking for the
sense of these notions in Arian theology,   trying to get out of
the Trinity College doctrine which preached relativism and
logically inconsistent trinity. One can think that for Newton as a
member of Trinity College at Cambridge University this situation
was a paradox \cite{ar}.
 }
 in his dramatic discussion with the founders of this
 doctrine\footnote{One of the
 founders of the Trinity theology Basil the Great revealed
 limits
 of the {\it consistent Aristotle's logic}
(see \cite{flo,mey}) which anticipated the famous G\"odel theorems
formulated in the 20th century for {\it formal arithmetical
systems}: ~i.  Any {\it consistent} description of a system is not
{\it
 complete},

 \hspace{1cm}ii.  Any {\it complete} description of a system is not
 {\it consistent}~\cite{god1}.

\noindent Here {\it consistent} means the absence of logical
contradictions. The theological deed of Basil the Great was the
positive constructive generalization  of the {\it consistent
Aristotle's logic}, in contrast with the negative G\"odel
theorems. Basil's generalization of {\it  Aristotle's logic}
 can be formulated as a continuation of the first
two  G\"odel theorems:

 \hspace{1cm}iii. Any {\it complete} description of a
system with logical contradictions becomes
 {\it consistent} in a new meaning, if there are simultaneously
  two (or more)
 descriptions of
 the system
connected by {\it relationships}, so that two opposite assertions
about the same system belong to two  their different {\it
descriptions}.
}.

\section{Newton's doctrine versus the ``Trinity'' one}

\subsection{Two ``descriptions'' of one relativistic object}

 %%%%%%%%%%%%%%%%%%%%%%%%%%%%%%%%%%endd
 David Hilbert's
formulation of GR\footnote{In the year of celebration of the 90th
anniversary of GR
 we can now distinguish Einstein's treatment of general coordinate
 transformations (who considered them as
   generalization of the frame transformations \cite{einsh}) from Hilbert's
   variation approach to GR \cite{H}, where the general coordinate
 transformations are considered as {\it diffeomorphisms} of the GR action
 and a {\it geometric
 interval}.
  There is
  an essential difference between the frame group
 of the Lorentz -- Poincar\'e-type  \cite{poi}
  leading to a set of
 initial data
 and the {\it diffeomorphism  group} of general coordinate transformations
  restricting these initial data by {\it constraints}. This difference
  was revealed by two  N\"other theorems  \cite{Noter}.
   The formulation
  of GR in terms of
 the Fock simplex \cite{fock29} defined as a diffeo-invariant Lorentz
 vector helps us  to separate  diffeomorphisms from transformations of
 frames of reference.} in a comoving
frame of reference can be an effective illustration of the
adequateness of the  Trinity doctrine to description of any
relativistic systems (particle, string, universe in GR).

 In order to demonstrate this adequateness, let us
 consider the problem of measurements of the life-time of
 an unstable  particle being in a relativistic train moving with velocity 200 thousands km/s.
 If a Driver  of the train measures its life-time 10 s, then  a Pointsman
 at a railway station measures 14 s (see Fig. 1).
 What is a correct life-time, 10 s or 14 s?

\begin{figure}[t]
\vspace{-1cm}
 \begin{center}
\includegraphics[width=0.6\textwidth,clip]{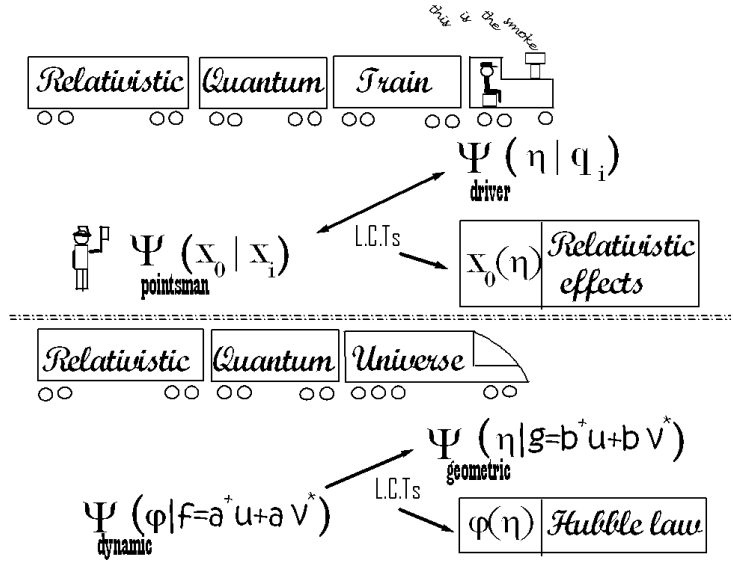}
\caption{{\small At the top of Fig.1 a relativistic train is
depicted with an unstable particle. The life-time of this particle
is measured by two observers, by a Pointsman and a Driver.
  Each of the observers has  his time:
 Pointsman - variable $X^0$, and Driver - geometrical interval
 $\eta$ (and his wave function).
 At the bottom of the figure there is an image of
 the universe where each observer has two
 sets of measurable quantities corresponding to two observers of the particle.
 To the Pointsman there corresponds a field set of measurable quantities
 (mass $\vh$ scaled by the cosmological scale factor and density of a
 number of particles), and the Driver --
  geometrical set of measurable quantities (time interval
  $\eta$  and initial data)~\cite{pvng6,pvng7,pvng8}.
} \label{fig2}}
\end{center}
\end{figure}

Two different times of the same particle
  is contradiction, in the Newton  doctrine\footnote{
  Resolving
 the Newton equation (the acceleration
  is equal to zero) one can find
 a  trajectory of the train, i.e. the
 time dependence of the coordinate of the train $X(t)=X_I+V_It$,
 where $V_I$ is a ratio of the momentum
 $P$ and a mass $m$.
The passenger can choose another frame with other
 data, but the time of a particle in all frames will be
unique and absolute, as it was postulated by Newton in his
mechanics with the  energy $E_{\rm Newton}(P)={P^2}/(2m)$. }.

Two different times of the same particle are described
  by two different  actions given in two different frames (comoving and rest),
  in the Einstein-type intermediate  doctrine\footnote{
 Lorentz, Poinc\'are, and  Einstein saw that electrodynamics
 was consistent with mechanics, where the train had the trajectory
 $X(X_0)=X_I+\frac{P}{E(P)}X_0$,
 and the energy $E(P)=\sqrt{m^2+P^2}$.
 This energy revealed and explained nuclear energetics.
 If an observer chooses another frame, the time $X_0$ converts
 into another one mixing with the coordinate
 by the kinematic Lorentz transformation
  %justified by the symmetry of
 in accordance with the Maxwell
 electrodynamics.
 This fact of changing time is one of relativistic effects
 meaning that each frame has its own time of a particle.
}, where
 the relation between these two times is only a kinematic one, but not
  a consequence of dynamic equations of motion.

 Two different times of the same particle and their relation
 as a consequence of dynamic equations of motion
 are  described by
 one Hilbert's geometro-dynamic action given in one concrete frame,
  where there is a
 bifurcation of the classical time into
 two different times (time-variable  measured by a Pointsman
  and time-interval measured by a Driver)
 of the same object.
   Isaac Newton could know in this   bifurcation the Trinity
   doctrine, because
  these two times belong to
  two different {\it descriptions} (here the dynamic
 and the geometric) of the object by the same action in the same frame\footnote{
 In order to give  the dynamic description of the relativistic
 relations, one should consider the time $X_0$ as one of the dynamic
 variables in the World space of events $[X_0|X_1]$ on equal
 footing with spatial coordinate $X=X_1$ and introduce one more
 time as a geometric measure $\eta$ of a trajectory
 $X_0(\eta)=\frac{P_0}{m}\eta,~X_1(\eta)=X_{1I}+\frac{P_1}{m}\eta$
 of a particle
 in the World space of events $[X_0|X_1]$. These equations are supplemented by
 the energy constraint $P_0^2-E^2=0,~ E=\pm\sqrt{P_1^2+m^2}$
  as an equation of the metric
 component of the Einstein interval expressed in terms of an unmeasurable
 ``coordinate time`` as an object of ``general coordinate transformation''
  in the one-dimensional  Riemannian space.
}.

 The Hilbert geometro-dynamics of a particle leads not only to Einstein-type
 theory with the  energy  explaining nuclear processes,
 but also predicts an antiparticle with a negative energy
  known as
 a Dirac positron and
reveals the historical pathway of quantum field theory  (the main
tool  of  high energy physics in the 20th century) because the
quantum field theory is nothing, but  primary and secondary
quantizations of the energy constraint,
  in order to remove a negative energy and to make a particle stable.
 However,
 the  quantum field theory uses only a time-variable
  measured by an external observer (i.e. a Pointsman)
 being out of a particle.

\subsection{Hilbert's Geometro-Dynamics of the Universe}

 In the Universe,
 an observer is simultaneously a Driver and a Pointsman.
 To the Pointsman there corresponds a field set of measurable quantities
  (the cosmological scale factor\footnote{
  In  GR,  the cosmic evolution is extracted by
 a scale transformation
${g}_{\mu\nu}=a^2\widetilde{g}_{\mu\nu}$ well known as the
cosmological perturbation theory \cite{lif,bard}. The GR action
 in
 terms of metrics $\widetilde{g}$ depends on only
 the running Planck $\vh(\eta)=aM_{\rm
 Planck}\sqrt{3/{8\pi}}$
mass scaling   all masses.}
  considered as a {\it time-like
  variable}  and densities of matter), and the Driver --
  geometrical set of measurable quantities ({\it
  time-interval}
   and initial data).
   In this case,
 geometro-dynamic relation between the {\it time-variable} and
 {\it time-interval}\footnote{
  The metric component
     $\sqrt{-\widetilde{g}}~\widetilde{g}^{00}=(\widetilde{N}_d)^{-1}$
     gives
 the conformal time of photons flying in the conformal
 flat space-time $\zeta_{(\pm)}=\int dx^0\langle N_d^{-1}\rangle^{-1}=
  \pm\int^{\vh_0}_{\vh} d\vh/\langle{(\widetilde{T}_0^0})^{1/2}\rangle$, where
  $\langle F\rangle=V_0^{-1}\int d^3x F$ is the averaging over finite
  volume $V_0=\int d^3x$, and $\widetilde{T}_0^0$ is the energy-momentum
  tensor component.} is just the famous
   Hubble law: the further a  star, the more
 redshift of the star photons. As the cosmological scale factor is
  the {\it time-variable}, its momentum
  is  the {\it energy of the
   Universe}\,\footnote{The energy constraint takes the form
  $P^2_\vh-E_\vh^2=0$, where $P_\vh$ is the scale factor canonical momentum
 and $E_\vh=2\int d^3x(\widetilde{T}_0^0)^{1/2}$ should be treated as
 the ``frame energy'' of the Universe like $E=mc^2$ in Special Relativity.}
 in the field space of events. In such the Einstein -- Hilbert GR
  the cosmic evolution is consistent with
  primary and secondary quantizations of the energy
  constraint\footnote{The primary quantization of the energy
  constraint $[\hat P^2_\vh-E_\vh^2]\Psi_L=0$ leads  to
  the unique  wave function $\Psi_L$
  of the
  collective cosmic motion.}
   justified
  by the experience of the high energy physics the 20th century.
 These quantizations  gave the
  description of the creation of the
 Universe and its geometric time and  matter from the physical
 vacuum\footnote{ The secondary quantization
 $\Psi_{\rm
L}=[{1}/{\sqrt{2E_\vh}}][A^++A^-]$ and Bogoliubov's
transformation:
 $ A^+=\alpha
 B^+\!+\!\beta^*B^-$ diagonalizes the equations of
 motion by the condensation of ``universes'' $<0|({i}/{2})[A^+A^+-A^-A^-]|0>=R$
 and describes   creation of a  ``number'' of
 universes
  $<0|A^+A^-|0>=N$
  from the stable Bogoliubov vacuum  $B^-|0>=0$.}
  \cite{pvng6,pvng7,pvng7a,pvng8b} in
  satisfactory agreement with observational
  data, if we chose
  the relative units \cite{pvng5}.
  %It was shown   that the Early
% Universe looked like  a factory of W-,Z- bosons created from the stable
% physical vacuum

 \begin{figure}[t]
\vspace{1cm}
 \begin{center}
 \includegraphics[width=0.65\textwidth,clip]{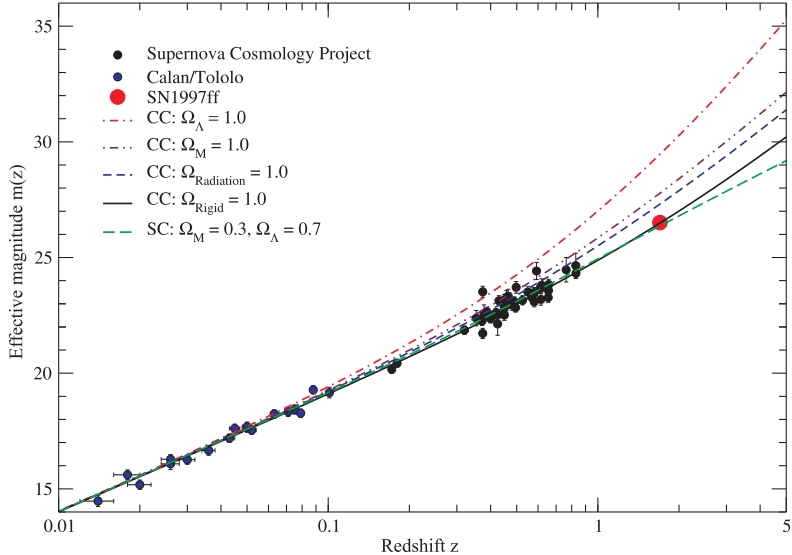}
\caption{ \small The Hubble diagram~\cite{pvng5} in cases of the
absolute units of standard cosmology (SC)  and the relative ones
of conformal cosmology (CC).
 The points include  42 high-redshift Type Ia
 supernovae~\protect\cite{SN2} and the reported
 farthest supernova SN1997ff~\protect\cite{SN1}. The best
fit to these data requires the dominance of the cosmological
constant density (70\%) and 25\% of the Cold Dark Matter density,
 in the case of the Friedmann  standard
cosmology, whereas for the relative units
 these data are consistent with  the dominance of density
of a free scalar field (85\%$\pm$ 10\%) with the square root
dependence of the scale factor on conformal time (that corresponds
to the stiff state) \cite{pvng5,Danilo}. \label{fig1}}
\end{center}
\end{figure}

 \subsection{Relative Units}

The cosmic dynamics of the Universe was revealed in General
Relativity by Alexander Friedmann \cite{f22} who kept in the
Einstein interval only a cosmological scale factor and resolved
equations of motion in this case. However, the interpretation of
 this cosmic dynamics as the expansion of the universe is possible
 only in the Newton doctrine of absolute units,
  if  we propose that our standards
  belong to the kingdom of {\it laws}.

 In accordance with the Trinity doctrine \cite{flo,mey},
 a man is free to choose his
 standard himself, because
 we can cognize  only
 a ratio of things. If a measurement is one of the tools of
 cognition, the Trinity doctrine means that
  we can measure any physical quantity only in units
  of another physical quantity accepted as a
  standard\footnote{Maxwell revealed that the description of results
   of experimental measurement of  electromagnetic phenomena
   by the field theory equations
 depends on the definition of measurable quantities in the theory
  and the choice of their measurement standard.
 In the introduction
  to his {``A Treatise on Electricity and Magnetism''} Maxwell wrote:
{\it ''The most important aspect
 of any phenomenon from  mathematical
 point of view  is that of a measurable quantity.
  I shall therefore consider electrical phenomena
  chiefly with a  view to their measurement,
 describing the methods of measurement, and
 defining the  standards on
  which they depend.''}%(J.C. Maxwell)
 \cite{Maxwell}.
 }.

 Defining a measurable interval of the length as the ratio
 of a Friedmann-like interval to the standard
 one that  also  belongs to  Friedmann-like intervals, one can see
 that the measurable interval of the length is not expanding,
 as it does not depend on the  cosmological scale factor.

 Really,  the scale factor
 disappears from the relative measurable interval, but not from the
 equations of motion. The equations of motion contain the
 Friedmann  cosmological scale factor as a measure of all  masses.
 It was shown~\cite{pvng5,pvng6,pvng7,pvng8} that
the relative units give a completely different physical picture of
the evolution of the universe than the absolute units of the
standard cosmology. The spectrum of photons emitted by atoms from
distant stars billion years ago remains unchanged during the
propagation and is determined by the mass of the constituents at
the moment of emission. When this spectrum is compared with the
spectrum of similar atoms on the Earth which, at the present time,
have larger masses, then a redshift is obtained.

The temperature history of the expanding universe
 copied in the relative units looks like the
history of evolution of masses of elementary particles in the cold
universe with a constant temperature of the cosmic microwave
background. The relative observable distance  loses the
cosmological scale factor $a$, in comparison with the absolute
one. Therefore, in this
 case, the observational redshift --
  coordinate-distance relation \cite{SN2,SN1}
  corresponds to the dominance of the stiff state
of free scalar field with the square root dependence of the scale
factor on conformal time   \cite{pvng5} (see also Fig. 2). Just
this time dependence of the scale factor on
 the measurable time (here -- conformal one) is used for description of
 the primordial nucleosynthesis \cite{three}.

\subsection{Electro-Weak Epoch  versus the Planck one}
Thus, the relative units doctrine  leads to a single stiff state
for
 all epochs including the creation of a quantum universe
 at the beginning.
In terms of the relative units
 the Planck mass loses  its  status as the fundamental parameter
  of the equations of motion and  becomes the present-day
  value of
 the running mass scale in contrast with the Inflationary Model
 \cite{bard,linde}
 based on the proposal   about the existence of the  Planck epoch
   at the beginning.
 The initial data of the creation  are determined by
 parameters of matter cosmologically created from the stable
 quantum
 vacuum  at the beginning of a universe.
 In the Standard
 Model,  W-,Z-vector bosons have maximal probability of this
 cosmological creation
 due to their mass singularity at
   the moment when their  Compton
   length  is close to the
 universe horizon defined by the primordial Hubble parameter.
 Therefore, the Universe was a factory of  W-,Z-vector bosons
 at the beginning. The observational data on CMB
 reflect the   parameters  of Standard Model of elementary particles
 (see Fig. 3,4) \cite{pvng7,pvng8}\footnote{Now it is a single possibility to explain the CMB
 observations  \cite{WMAP} in the
  framework of GR and SM, as it was shown in  \cite{pvng7a,pvng8b}
 the standard cosmological perturbation theory \cite{lif,bard}
 applied for  analysis of the CMB observations \cite{WMAP}
  in the Inflationary Model \cite{linde}
  has  a different number of variables
  than the Einstein  General Relativity  \cite{einsh}.}.

\begin{figure}[t]
 \centerline{\includegraphics[width=3in]{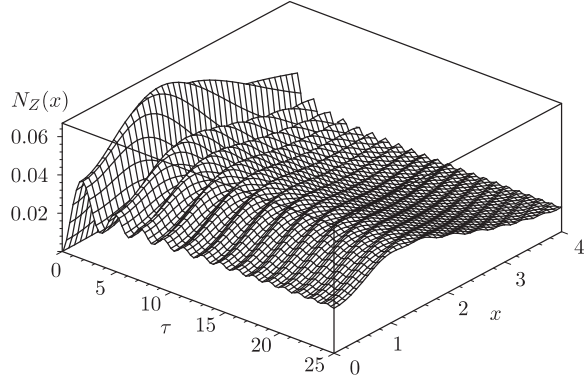}}
 \caption[]{\small The surface shows the distribution function of creation of
 longitudinal ($N_Z(x)$) components of the W-bosons in units of
   the dimensionless time ($\tau= 2\eta H_I$) and
the dimensionless momentum ($x = q/M_I$) with the constant
conformal temperature
 $T_c\sim (M_I^2H_I)^{1/3}=(M_{\rm W}^2H_0)^{1/3}\sim 3 K$. These bosons created
 from vacuum at   the moment when   their  Compton   length
 defined by the inverse mass
 $M^{-1}_{\rm I}=(a_{\rm I} M_{\rm W})^{-1}$ is close to the
 universe horizon defined in the
 stiff state as
 $H_{\rm I}^{-1}=a^2_{\rm I} (H_{0})^{-1}$.
 Equating these quantities $M_{\rm I}=H_{\rm I}$
 one can estimate the initial data of the scale factor
 $a_{\rm I}^2=(H_0/M_{\rm W})^{2/3}=10^{-29}$ and the Hubble parameter
 $H_{\rm I}=10^{29}H_0\sim 1~{\rm mm}^{-1}\sim 3 K$.
 CMB radiation is described as the product of decay of
primordial  W-,Z- bosons during
 the time-life $\eta_L\sim (2/\alpha_W)^{2/3}(T_c)^{-1}$ expressed  in terms of
the
  Weinberg coupling constant $\alpha_W=\alpha_{\rm \tiny QED}/\sin^2\theta_{\rm Weinberg}\sim
 0.03$
   \cite{pvng7,pvng8}.
 The primordial mesons before
 their decays polarize the Dirac fermion vacuum and give the
 baryon asymmetry frozen by the CP -- violation
 so that for billion photons there is only one baryon,
  and relative contributions of the baryon matter and radiation
  are in agreement with observational data:
 $\Omega_b \sim \alpha_W\sim
 0.03$, and $\Omega_R\sim 10^{-5}\div 10^{-4}$~\cite{pvng8}.
The equations of the longitudinal vector bosons
 in SM, in this case,
 describe the ``power primordial spectrum'' of the CMB radiation.
} \label{fig3}
\end{figure}
\begin{figure}[]
%\figurenum{1} \epsscale{0.6} {\plotone{f1j.eps}}
 \centerline{%\rotatebox{90}
 \includegraphics[width=0.55 \textwidth,height=0.71\textwidth,angle=90]{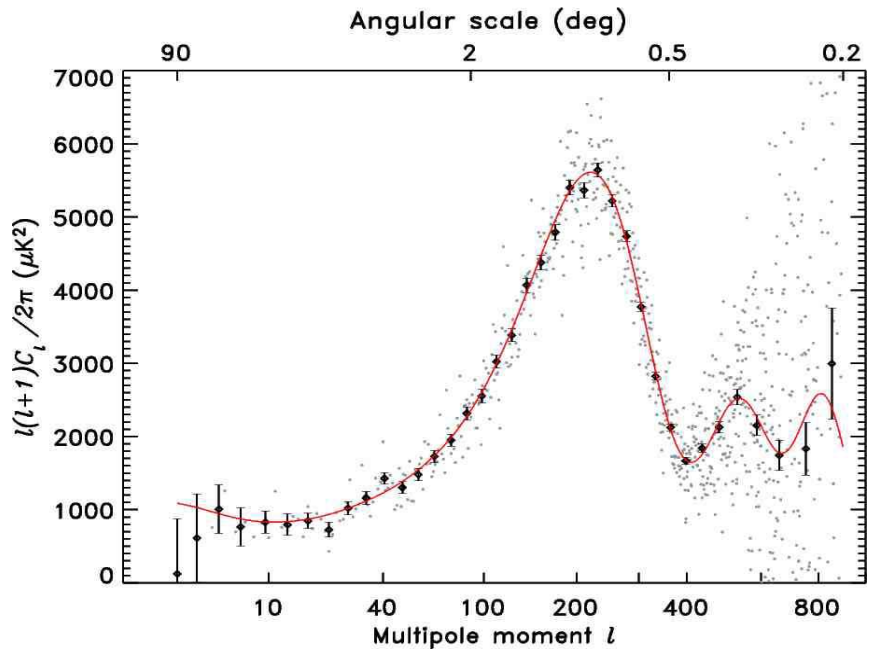}}
 \caption[]{\small The curve  shows power law of the
 temperature angular power spectrum with first peak $l\sim 210$ \cite{WMAP}.
 The parameters of CMB can reflect parameters of the SM of elementary
 particles, in particular the Weinberg coupling constant
 \cite{pvng7,pvng8,pvng8b}.}
\end{figure}

 The geometro-dynamic  formulation
 of General Relativity  \cite{dir,Dir,H,fock29} supplemented by
 the cosmological scale factor as a collective time-like variable
  in a comoving frame and ``relative'' units \cite{pvng8,pvng7a,pvng8b}
  corresponds rather  the Trinity
  doctrine than the Newton one. This
  geometro-dynamics
 allows us not only to restore the historical pathway of physical
 description of the Universe evolution
 but also to use experience of quantum field theory
formulation verified by the high energy physics experiments. We
gave here  a set of numerous arguments in favor that
 this quantum unified theory (GR \& SM \& an additional scalar field)
  can be a theoretical basis
 of  the further detailed investigation of astrophysical
 observational data including
  CMB fluctuations as one of the highlights of
 present-day cosmology with far-reaching implications and more
 precise observations are planned for the near future.

\section{Design and Purpose in the Universe}

 Thus, the papers of the founders of  modern
 physics \cite{dir,Dir,H,einsh,fock29}
  allow us to restore
 the historical pathway of description of the Universe evolution
 in the comoving frame   with the definite initial data in the unified theory
 like the description of the trajectory of a relativistic
 particle in SR after its quantum creation from  stable
 vacuum. This description reveals the trinity of times in GR:
 i) the unmeasurable  coordinate time in  Hilbert's action as
 an object of general coordinate transformations (i.e., diffeomorphisms),
 ii) the diffeo-invariant geometric time-interval measured by our watch,
 and iii) the time-like variable as the cosmological scale factor
 measured by astrophysical observations and considered as a measure
 of the Universe evolution in the field space of events,
 where the Universe was created \cite{pvng1} -- \cite{pvng8b}.

 One can say that modern physics goes beyond the bounds of
 the Newton Arian doctrine accepted by modern theoreticians
 \cite{mo}
 who   have a possibility to  continue
 the Newton dialogue  with the founders of the Trinity doctrine
 \cite{flo,mey}.
 Who is a man in the Universe?
 He is a passenger who lost his rest frame in the ``frame free'' method,
 who uses only absolute units and parameters of the ''natural'' laws
 including the ``absolute Planck mass'',
 and cannot determine unambiguously the
 energy of the Universe?
 Or he is a Driver, who knows his frame and place,
  chooses his
 relative units,  measures the initial data,
 introduces new concepts and changes his
 logics as one of the tools of his cognition,  who is able to determine
 the Universe energetics, describes
  its creation, and is responsible for its further fate, as he has his
  free ``will'', in order to take part
  in further creation of the measurable world and
  to have an eternal life of his unmeasurable soul in the
 kingdom of His {will}\,?

%{\bf Acknowledgements}\\
% The author is grateful to Profs. B.~Barbashov, Z. Oziewicz,
%  and V.~Priezzhev for fruitful discussions.

%\newpage

%\noindent

\newpage

\section{Discussion at the 42nd  Karpacz Winter School of
Theoretical Physics}

 I am very grateful  to Profs. A. Borowiec and M. Francaviglia
 for the hospitality at the 42nd  Karpacz Winter School of Theoretical
 Physics on Current Mathematical Topics in Gravitation and Cosmology, L\c adek,
 Poland, 6-11.02.2006
 {\small (http://www.ift.uni.wroc.pl/karp42/)}.
 At the School there were very interesting and
 sharp debates on energy in General relativity
 (GR), its variables, and    a reference frame of the Universe
 creation.
 I list here some questions and my responses.

 \subsection*{FRAME?}

  {\bf Question by
 Prof.  Zbigniew Oziewicz} (Universidad Nacional Autonoma de M\'exico):

 What does ``the concrete reference frame of the Universe creation'' mean?
 A frame is the coordinate basis, and  physical results should not
 depend on any ``basis''.

\vspace{.31cm}

 \noindent {\bf Response (V.P.)}:

 Let me recall that the introduction of the physical concept
 of a ``reference frame'' as a three-dimensional coordinate basis with a watch
  was really stimulated
  by the great Polish astronomer  M. Kopernik who
 proposed in 1543 the heliocentric  ``frame''  as an alternative to
 the Ptolemaeus ``frame''  connected with the Earth observer.

 Thus, there
 appeared a definition of  ``physics'' as a science about observations
  (later --
 measurements) of objects being in the ``comoving frame''
  of their center of masses by devices
  associated with the ``rest frame''  of an observer.
 In the context  this definition
 Kopernik's ``relativity'' is nothing, but
 different epicycles ``relative'' to different frames.

 Galilei in his ``Dialog ...'' in 1632 converted these
 two  frames into a set of inertial frames  moving with  constant
 velocities.
  He identified one of the inertial frames with   a ship,
  where a motion of the measurable object  ``relative'' to
  a moving device is described by the difference of their
  two velocities. In this case, you are right,
  different motions of the ship do not change indications
  of the devices, i.e. initial data. I meant only motions of
  a comoving frame
  relative to a rest frame, i.e. transformations of the initial data, because
 we now have a similar situation   with
 the Cosmic Microwave Background (CMB) measured by the SkyLab
 ``Hubble'' devices being in the ``Ptolemaeus (rest) frame'' moving
 ``relative'' to CMB with the velocity 400 km/s to Leo.
 If we  move the SkyLab ``Hubble'' together with CMB
 (i.e. make the Lorentz transformations changing
  the initial data),
 the dipole components
  of CMB temperature  disappears (see \cite{pvng8a}).
 Thus, in accordance with
 Kopernik's ``relativity'' principle, the Universe as
 any ``observable'' object   has
 its ``comoving frame''  marked by CMB.

\vspace{.31cm}

\noindent {\bf Question by
 Prof.  Zbigniew Oziewicz}: %(Universidad Nacional Autonoma de M\'exico):

 What is your opinion about the relation of the Lorentz group
  to electrodynamics?

\vspace{.31cm}

 \noindent {\bf Response  (V.P.)}:

 The Faradey -- Maxwell electrodynamics contained the hidden
 symmetries revealed later by Poincare (known as the Lorentz group of
  transformations of the initial data) and
  by Weyl (known as gauge symmetry leading to constraints of initial data).
  Einstein's ``Special Relativity'' is  his generalization of mechanics
  based on the Lorentz group of frame transformations leading to
  different times ``relative'' to
   different frames and  energy $E=mc^2$ in the comoving frame.

 Einstein's ``General Relativity'' was treated by him as
 generalization of the Lorentz frame
  transformations; whereas Hilbert in his ``Foundations'' \cite{H}
  considered GR as a ``gauge'' theory with a set of constraints
  revealed by the Hilbert theorem known as the second N\"other one.
  The Lorentz transformations were introduced in GR by means of
   the Fock tetrads \cite{fock29}.

\subsection*{VARIABLES?}
 Problem of the choice of variables in different
 theories of gravitation was considered in Lectures by
 Profs. {\bf M. Francaviglia}, {\bf G. Allemandi}
 (both from  University of Torino) and
{\bf  L. Soko\l owski} (Jagiellonian University, Krak\'ow). In all
these lectures the remarkable fact
 was noted. All metric theories of gravitations are
 connected by general conformal transformations of their variables and
 they are mathematically equivalent.
 The questions appeared: What variables can be considered as
 measurable quantities? and What is the criterion of the choice
 of a true gravitation theory among the mathematically equivalent
 ones?

\vspace{.51cm}

 \noindent {\bf Response (V.P.)}

 In the context of the results presented in these Lectures
  the action of Einstein's GR is mathematically equivalent to the
  negative action of the
  conformal invariant theory of a  scalar field
  (called ``dilaton'') \cite{pvng2,pvng5}.
 The latter does not contain any dimensional fundamental
 parameter of the type of the Planck mass.
 If this action is supplemented by the SM one, where the Higgs mass is
 replaced by the ``dilaton'' field, we get the unified theory,
 where the scale (and, therefore, conformal) symmetry can be broken by the
 initial data at the beginning of the Universe provided that
 the present day value of the ``dilaton'' field is equal to the Planck mass.
 We see that the idea of conformal symmetry of the World developed
 by Richard R\c aczka (see cites in \cite{pvng2}) excludes the
 Planck epoch at the beginning of the Universe.
 Conformal symmetry distinguishes the
 Lichnerowicz {``conformal variables''}
 $g_{\mu\nu}^{(L)}=|g^{(3)}|^{-1/3}g_{\mu\nu}$
  constructed  with the help of the spatial metric determinant
  and used by Dirac in his Hamiltonian approach \cite{dir}.
  Identification of the Lichnerowicz {``conformal variables''}
  with the measurable quantities leads to Conformal Cosmology
  with varying masses (instead of the expanding Universe)
 where Supernova Data are described by the ordinary
 homogeneous free scalar field in the stiff state
 without Cosmological Constant \cite{pvng5}.

\subsection*{ENERGY?}

  The  heaviest debates at the Winter School  were
 about the choice of the energy in General Relativity
 and cosmology.
 Profs. {\bf J. Garecki} (University of Szczecin),
 {\bf L. Lusanna} (INFN, Firence),
 {\bf  L. Soko\l owski}, and {\bf M. Dyrda}
  (both from Jagiellonian University, Krak\'ow)  in their talks
   used the conventional  definition of the ADM-type energy
   relative to an external
  observer in asymptotically flat infinite spacetime.
 Profs. {\bf M. Francaviglia} (University of Torino), {\bf J.
 Lukierski} (ITF, University of Wroc\l aw) and other participants
   insisted on
 that just this
 definition could not be applied in cosmology
 because an internal observer
  has a possibility to observe only
    finite spacetime of the Universe. The  illustration
    of this discussion was  the remarkable
      film by {\bf Lorenzi Marcella} (Cosenza, Italy)
     about the Einstein energy  $E=mc^2$.

\vspace{.51cm}

 %:My talk One can suppose

%describing
% the collective motion of the observable relativistic object,

 \noindent{\bf Response  (V.P.)}

  In the Hilbert-type formulation
     of Special Relativity (SR), $E=mc^2$  is considered as a solution of
     the energy constraint  with respect to the canonical momentum
     of the time-like variable.
     Just the  GR generalization  of this SR energy $E=mc^2$
 to  the Universe in finite space
 was the topic of my talk  \cite{pvng1,pvng7a,pvng8a,pvng8b},
 where the collective time-like variable %$\vh$
 is identified with the
 homogeneous scale factor $a$
 multiplied by the Planck mass. It is a varying Planck mass
 $\vh=a M_{\rm Pl.}\sqrt{3/(8\pi)}$ considered as the
 homogeneous ``dilaton``  \underline{degree of freedom }
 in conformal version of GR.
 The Einstein-type energy in GR (as a solution of
 the energy constraint) depends on the varying mass $\vh$.
 Just  this dependence
 leads to the \underline{freedom} to
 create  the Universe and its matter from the vacuum
  in the comoving frame with the
 initial data $\vh_I^2/\vh_0^2 =a^2_I=H_0/H_I\sim 10^{-29}$ \cite{pvng7,pvng8}.

  \vspace{.51cm}

  \noindent {\bf Question by Prof.  M. Szyd\l owski} (Jagiellonian University, Krak\'ow)

  What does creation of universes from the vacuum mean?

 \vspace{.31cm}

 \noindent {\bf Response (V.P.):}

  QFT of universes is formulated  by analogy
  with  the  QFT of particles \cite{Dir} as
 the primary and the secondary quantizations of
 the  energy constraint,
 in the context of Hilbert's variational principle \cite{H}, in order to remove
 the negative Einstein-type energy
  and
 to make stable the quantum system\footnote{The existence of a stable vacuum
 as the state with minimal energy can be considered as the quantum
 analog of the Boltzmann H-theorem about the existence of
 stable temperature.}.
 We need only to understand: What does QFT mean for the ``varying masses''?
 But this problem was solved  at the end of the 60s
 for particles by Chernikov and  Parker
  with the help of  the Bogoliubov
 transformation (1946)  and
 for universes where the role of the ``mass'' is played by the
 cosmological energy density
  in 2006 \cite{pvng7a,pvng8a,pvng8b}.

 %\vspace{1cm}
\subsection*{DEGREES OF FREEDOM?}

 \vspace{.41cm}

 \noindent {\bf Question by Prof. S. Odintsov} (Barselona University \& Tomsk
 University)

  What is the status of the conformal sector of Quantum Gravity
  considered in the papers

    I. Antoniadis, E. Mottola, PRD  V. 45, 1992, p.2013;

   S.D. Odintsov, Zeit.fur Physik  V. C54, 1992, p.531;

   I. Antoniadis and S.D. Odintsov, Phys.Lett. B343, 1995, p.76 ?

 \vspace{.51cm}

 \noindent {\bf Response (V.P.):}

  These papers were devoted to GR in infinite spacetime without any
 supposition about a stable vacuum, negative energy,  initial data,
 time-variable, units~of measurement, and separation of all
 variables and metric components into degrees of \underline{freedom} with
 gauge-invariant initial data violating the frame symmetry and
 potentials without initial data, and they explained nothing.

 In particular, the kinetic energy density  of the conformal sector is
 negative. Therefore,  Dirac  (in his Hamiltonian
 approach to GR formulated in the ``comoving frame'' \cite{dir})
  identified all scalar components of the conformal sector with
 potentials.
  This identification  is in agreement with the Schwarzschild solution,
  but not with cosmological evolution that appears in the
  conformal sector,
  in the case of finite spacetime,
  as the homogeneous zero mode of the momentum constraint,
  i.e the scale factor.
  Lifshits \cite{lif} treated this cosmological scale
 factor as an additional homogeneous variable.
 However, this treatment  leads to double counting of the homogeneous variable
 and  loses a possibility
 to formulate a Hamiltonian approach
  together with quantization and the freedom
 to create the Universe from the vacuum \cite{pvng8a}.

 Dirac had no  homogeneous degree of freedom (with Hamiltonian
 and without cosmology), and Lifshits   introduced  two  (with cosmology
 and without Hamiltonian).

 In the finite volume Hamiltonian approach to GR \cite{pvng8a,pvng8b}
 the scale factor is introduced as the spatial averaging
  the spatial determinant logarithm\footnote{In this case
   the Newton potential is defined
 in the class of function with nonzero Fourier harmonics, i.e.,
 the  spatial averaging of this potential is equal to zero. This fact
  reveals a difference of the ADM-type  energy based on this potential
  from  the cosmological energy density  given in finite space
  by means of the  spatial averaging.}. The negative sign of the energy density
  of the cosmological scale factor  shows us that it is
  the evolution parameter, and the constraint value of
  its canonical momentum is the constrained energy like
  the Einstein one $E=mc^2$ in SR.
  %It is easy to convince that QFT of universe describes
%  the creation of the Universe like QFT of particles describes
%  the creation of particles.

 %The  cosmological perturbation theory \cite{lif}
% applied in the Inflationary Model \cite{bard}
% to  explain CMB spectrum   contradicts
% to the Hamiltonian approach.

\subsection*{COSMOLOGY?}
 The set of Lectures by {\bf Salvatore Capozzielo} (University of Napoli)
 was devoted to the analysis of
 ``an impressive amount of different astrophysical data''
 in the class of the $\Lambda$ - Dark Matter  cosmological models
 with the present-day
 accelerating expansion, where the nature of the $\Lambda$ dark energy
  is
 still unknown. In this class of models, all principal cosmological problems
 (homogeneity, flatness, horizon, etc)
  are solved by the inflation mechanism \cite{linde}.
 The main  test of Inflationary Model is the derivation of ``CMB primordial
 power spectrum'' \cite{bard} based on the cosmological
 perturbation theory \cite{lif}.
 There were a lot questions at the School: What is difference of the
 finite volume Hamiltonian GR   \cite{pvng8b}
 from the Inflationary
 Model  \cite{bard,linde}?

\vspace{.51cm}

 \noindent {\bf Response (V.P.)}

 The main differences between the finite volume
 Dirac Hamiltonian approach to GR  \cite{pvng8b}
 and the Inflationary Model (IM) are the following.

 1. In the finite volume GR  the observational  quantities of
 ``the cosmological scale factor''
 and the ``conformal time'' are identified with the \underline{spatial volume averaging}
  the metric scalar components, so that their
 cosmological perturbations \underline{do not contain the zero Fourier
 harmonics}, in  contrast to the Lifshits cosmological perturbation
 theory \cite{lif} used in IM \cite{bard,linde}, where
  the ``conformal time''
 is determined by the physically unattainable procedure of
 synchronization of watches in the whole Universe, and
 ``the cosmological scale factor'' is introduced as an additional
 variable that
  increases the number of variables.

 2. In the finite volume Hamiltonian GR
 the Dirac principle of  stability of
 a relativistic quantum system is used in contrast with
 IM.
 The Dirac ``stability'' means i) the zero canonical momentum fields
 are gauge-invariant potentials, ii) existence of
 the vacuum as a state with a
 minimal constrained energy, iii) the zero momentum
 of the local volume element  removing its negative energy density
 and leading to a nonzero shift-vector $N^i$ of the coordinate origin
  with
 spacial oscillations\footnote{The test of the finite
  volume Dirac Hamiltonian GR is the
 large-scale structure of the Universe (see Fig. 5).}
  and the potential scalar perturbations, whereas the
  Lifshits-type kinetic scalar perturbations
 explaining CMB spectrum in IM disappear.

3. The  CMB spectrum can be reproduced, if  the absolute units of
 length in
 IM  are replaced by the relative units  for
 which SN data are compatible with
 the state of Dark Energy as a free scalar field \cite{pvng7}.
 In this case,  the Early Universe is a factory
 of W-,Z- bosons \cite{pvng8} the equations of which reproduce
  the Lifshits-type kinetic scalar perturbations
  explaining CMB spectrum in IM. This  factory gave us all
  matter and CMB as the decay
 product of the primordial bosons created from
 the stable ``vacuum'' after the creation of the Universe
 together with its coordinate frame and  time interval known here
  as the Kopernik  comoving frame.

  In  contrast to the IM in which the cosmic evolution
  begins with
  the Planck epoch and the Planck mass is treated as
  a fundamental parameter of the ``law'', in the Hamiltonian
  GR the Planck epoch
  becomes the present-day one and the Planck mass is treated as
  the law \underline{free} datum in the comoving frame of the Universe.

\subsection*{RESUME}

 It is useful to remember that
 Newton in his Trinity College strongly distinguished the {\it kingdom of laws}
 from the {\it kingdom of \underline{freedom} (of will)}. Newton knew
 that any {\it law} does not depend on {\it will}, but application
 of the {\it law} does. The creation of the Universe is
 the application of the {\it law}. The instruction of this  application
 can present here as the historical pathway of physics:
 from empiric facts to laws and
 from {\it laws to the grace} (John 1:17) of \underline{freedom}
  to create the Universe
 from "nothing" (Newton called this  history the {\it shoulders of giants}):

%\vspace{.21cm}

 Ptolemaeus' \underline{rest frame} (RF),

 \vspace{.21cm}

 Kopernik's \underline{comoving frame} (CF),

 \vspace{.21cm}

 Galilei's  \underline{inertial frames} as a set of
 all possible \underline{initial data} given  without the law (i.e. free),

 \vspace{.21cm}

 Newton's  \underline{differential law}  invariant with respect to
 transformations of \underline{the law free initial data},

 \vspace{.21cm}

 Faraday's program: the \underline{field nature} of matter,
 and \underline{unification} of forces of Nature,

 \vspace{.21cm}

 Maxwell's  equations with hidden \underline{Lorentz's frame
 symmetry} (Poincare) and \underline{gauge} one (Weyl),

 \vspace{.21cm}

 Einstein's \underline{Special Relativity} (SR) as
 \underline{different times for different frames} (RF \& CF),

 \vspace{.21cm}

 Einstein's \underline{General Relativity} (GR)  as
 generalization of the Lorentz frame group?,

 \vspace{.21cm}

 Hilbert's \underline{variational principle} for  GR, where Einstein's
 generalization becomes gauge group,

 \vspace{.21cm}

 Fock's \underline{tetrad} as the separation of the Lorentz frame transformations from
 the  gauge ones,

 \vspace{.21cm}

 Friedmann revelation of the \underline{cosmological scale factor as
 the collective  motion} in GR,

 \vspace{.21cm}

 Einstein's $E_{\pm}=\pm mc^2$ in Hilbert's constrained
 version of SR supplemented by the \underline{interval},

 \vspace{.21cm}

 Dirac QFT for particles as \underline{the primary and secondary quantization} of the energy
 constraint,

 \vspace{.21cm}

Lichnerowicz \underline{variables}
$g_{\mu\nu}^{(L)}=|g^{(3)}|^{-1/3}g_{\mu\nu}$
 distinguished by the spatial metric determinant,

\vspace{.21cm}

  Dirac - ADM \underline{Hamiltonian approach} to GR in terms
  of radiation-type Lichnerowicz \underline{variables},

\vspace{.21cm}

 Zel'manov's global gauge symmetry of the \underline{Hamiltonian GR}
 as foundation of cosmic evolution,

\vspace{.21cm}

 Finite volume Dirac \underline{Hamiltonian approach} to  GR with
 the \underline{cosmic  motion and its energy}

\vspace{.21cm}

 $E_\vh=2\int d^3x^{(L)}\sqrt{T^{0(L)}_0}$ as the canonical  momentum
 of $\vh=a M_{\rm Pl.}\sqrt{3/(8\pi)}$ \cite{pvng8b},

\vspace{.21cm}

Hubble law in the\underline{ conformal units} as the spatial
averaging the Lichnerowicz interval,

\vspace{.21cm}

Primary and secondary quantization of the energy constraint by
\underline{Bogoliubov's transformation},

\vspace{.21cm}

 Dirac -- Schwinger ``fundamental operator quantization'' \cite{sch2} of
 Weinberg -- Salam -- Glashow \underline{Standard Model} in the Kopernik-type
 comoving frame of the Universe, where all field components
  are separated into ``potentials'' (without initial
 data) and ``degrees of freedom'' (of initial data) with different
 poles of their propagators.

 Recall that Faddeev proved the theorem of
 equivalence of ``fundamental operator quantization'' \cite{sch2,pvn2}
  with the frame free ``Lorentz gauge
 formulation'' (where all field components are considered
  as ``degrees of freedom'' on equal footing)
  only for the scattering amplitudes. Nobody proved
  even in QED that instantaneous atoms and molecules formed by the
  Coulomb  potential
  can be obtained by the  Faddeev-Popov heuristic
  quantization \cite{fp1} in the frame free Lorentz gauge where
  all propagators have only the light cone singularities. To his
   great surprise a contemporary theoretician can
  know that  Schwinger
 ``{\it  rejected all Lorentz gauge formulations
as unsuited to the role of providing the fundamental operator
quantization}'' \cite{sch2}.
 Thereupon, it is worth  emphasizing that Schwinger following Dirac
  postulated the higher
 priority of the quantum principles in
 comparison with relativistic ones \cite{dir,Dir}\footnote{
 Modern M-theories with D=26, D=10,11 \cite{mo} are consequences of
 the ``scattering branch'' of the Faddeev -- Popov-type heuristic quantization
 \cite{fp1}.
 The fundamental operator quantization of a relativistic
 string leads to R\"ohrlich spectrum without the Virasoro algebra \cite{pvng7a}.
In any case, the SN data \cite{SN1,SN2} in relative units
compatible with
 Conformal Dark Energy  is described by a scalar field,
  so that the  scalar sector of the sum of GR and SM
 can be  the conformal
 D=6 brane in the 4th-dimensional spacetime.
}.

\begin{figure}[t]
\vspace{1cm}
 \begin{center}
 \includegraphics[width=0.6\textwidth,clip]{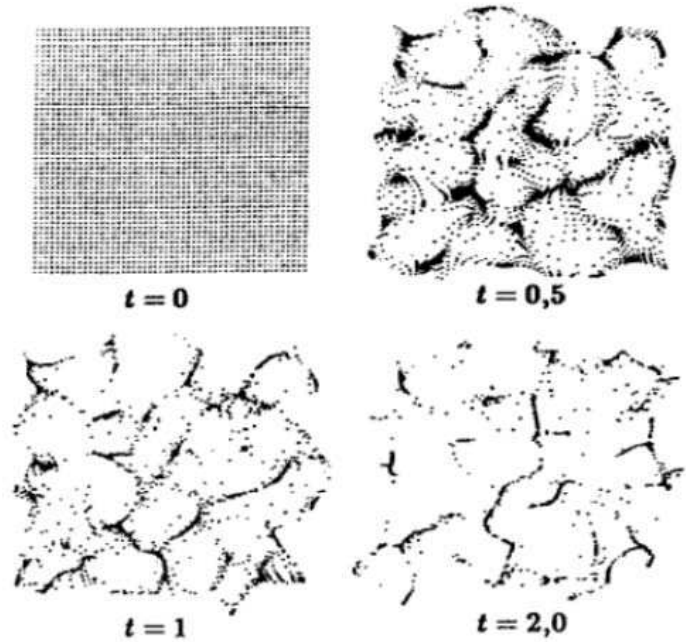}
\caption{
 \label{fig5} {\small
 The diffusion of
 a system of particles moving in the space
 $ds^2=d\eta^2-(dx^i+N^id\eta)^2$ with periodic shift
 vector $N^i$ and zero momenta could be understood from an analysis
 of the following system of O.D.E.  $dx^i/d\eta=N^i$ considered for the
 two-dimensional case.
 For the system we reproduce the results of numerical
 simulations given by
    Klyatskin V.I. ~``Stochastic Equations'' M.
 Fizmatlit, 2001, where $t=\eta m_{(-)}$, $
\overline{m}_{(-)}=H_0\sqrt{{6}[\Omega_{\rm
R}(z+1)^2+({9}/{2})\Omega_{\rm
  Matter}(z+1)]/7}$} \cite{pvng8b}.
The  size $\sim 130\, {\rm Mpc}$
  of spatial oscillations of matter is in agreement
  with value of radiation-type density
  $\Omega_R\sim 3\cdot 10^{-3}$ at
  the time of diffusion $z_r\sim 1100$ \cite{pvng8b}.
  }
\end{center}
\end{figure}

% Piechocki W\l odzimierz (Warszawa, Poland)
%

\end{document}